\documentclass[twocolumn]{aastex631}

\newcommand{\target}{MM\,J1545}

\received{January 1, 2024}
\revised{January 1, 2024}
\accepted{\today}
\shorttitle{Lensed SMG behind Lupus-I Molecular Cloud}
\shortauthors{Tamura et al.}
\begin{document}

\title{Large molecular and dust reservoir of a gravitationally-lensed submillimeter galaxy behind the Lupus-I molecular cloud}

\correspondingauthor{Yoichi Tamura}
\email{ytamura@nagoya-u.jp}

\author[0000-0003-4807-8117]{Yoichi Tamura}
\affiliation{Department of Physics, Graduate School of Science, Nagoya University, Furo, Chikusa, Nagoya, Aichi 464-8602, Japan}

\author[0000-0002-9695-6183]{Akio Taniguchi}
\affiliation{Department of Physics, Graduate School of Science, Nagoya University, Furo, Chikusa, Nagoya, Aichi 464-8602, Japan}
\affiliation{Kitami Institute of Technology, 165 Koen-cho, Kitami, Hokkaido 090-8507, Japan}

%% - alphabetical

\author[0000-0002-5268-2221]{Tom J.~L.~C.\ Bakx}
\affiliation{Department of Space, Earth, \& Environment, Chalmers University of Technology, Chalmersplatsen 4, Gothenburg SE-412 96, Sweden}

\author[0000-0003-4518-407X]{Itziar De~Gregorio-Monsalvo}
%%\affiliation{Joint ALMA Observatory}
\affiliation{European Southern Observatory, Alonso de Cordova 3107, Casilla 19, Vitacura, Santiago, Chile}

\author[0000-0001-8083-5814]{Masato Hagimoto}
\affiliation{Department of Physics, Graduate School of Science, Nagoya University, Furo, Chikusa, Nagoya, Aichi 464-8602, Japan}

\author[0000-0000-0000-0000]{Soh Ikarashi}
\affiliation{Fukuoka Institute of Technology, 3-30-1 Wajiro-higashi, Higashi-ku, Fukuoka 811-0925, Japan}
\affiliation{Department of Physics, General Studies, College of Engineering, Nihon University, 1 Nakagawara, Tokusada, Tamuramachi, Koriyama, Fukushima, 963-8642, Japan}
\affiliation{National Astronomical Observatory of Japan, 2-21-1 Osawa, Mitaka, Tokyo 181-8588, Japan}

\author[0000-0002-8049-7525]{Ryohei Kawabe}
\affiliation{National Astronomical Observatory of Japan, 2-21-1 Osawa, Mitaka, Tokyo 181-8588, Japan}
\affiliation{The Graduate University for Advanced Studies (SOKENDAI), Shonan Village, Hayama, Kanagawa 240-0193, Japan}

\author[0000-0002-4052-2394]{Kotaro Kohno}
\affiliation{Institute of Astronomy, Graduate School of Science, The University of Tokyo, 2-21-1 Osawa, Mitaka, Tokyo 181-0015, Japan}
\affiliation{Research Center for the Early Universe, Graduate School of Science, The University of Tokyo, 7-3-1 Hongo, Bunkyo-ku, Tokyo 113-0033, Japan}

\author[0000-0002-6939-0372]{Kouichiro Nakanishi}
\affiliation{National Astronomical Observatory of Japan, 2-21-1 Osawa, Mitaka, Tokyo 181-8588, Japan}
\affiliation{The Graduate University for Advanced Studies (SOKENDAI), Shonan Village, Hayama, Kanagawa 240-0193, Japan}

\author[0000-0002-4124-797X]{Tatsuya Takekoshi}
\affiliation{Kitami Institute of Technology, 165 Koen-cho, Kitami, Hokkaido 090-8507, Japan}

\author[0000-0001-9368-3143]{Yoshito Shimajiri}
\affiliation{Kyushu Kyoritsu University, 1-8 Jiyugaoka, Yahatanishi-ku, Kitakyushu, Fukuoka 807-0858, Japan}

\author[0000-0002-6034-2892]{Takashi Tsukagoshi}
\affiliation{Faculty of Engineering, Ashikaga University, 268-1 Ohmae-cho, Ashikaga, Tochigi 326-8558, Japan}

%\author[0000-0003-1937-0573]{Hideki Umehata}
%\affiliation{Institute for Advanced Research, Nagoya University, Furo, Chikusa, Nagoya, Aichi 464-8601, Japan}
%\affiliation{Department of Physics, Graduate School of Science, Nagoya University, Chikusa-ku, Nagoya, Aichi 464-8602, Japan}

%% - no response, need to confirm.

\author[0000-0001-6469-8725]{Bunyo Hatsukade}
\affiliation{National Astronomical Observatory of Japan, 2-21-1 Osawa, Mitaka, Tokyo 181-8588, Japan}
\affiliation{The Graduate University for Advanced Studies (SOKENDAI), Shonan Village, Hayama, Kanagawa 240-0193, Japan}

\author[0000-0002-2364-0823]{Daisuke Iono}
\affiliation{National Astronomical Observatory of Japan, 2-21-1 Osawa, Mitaka, Tokyo 181-8588, Japan}
\affiliation{The Graduate University for Advanced Studies (SOKENDAI), Shonan Village, Hayama, Kanagawa 240-0193, Japan}

\author[0000-0000-0000-0000]{Hideo Matsuhara}
\affiliation{Institute of Space and Astronautical Science, Japan Aerospace Exploration Agency, 3-1-1 Yoshinodai, Chuo-ku, Sagamihara, Kanagawa 252-5210, Japan}
\affiliation{The Graduate University for Advanced Studies (SOKENDAI), Shonan Village, Hayama, Kanagawa 240-0193, Japan}

\author[0000-0003-1549-6435]{Kazuya Saigo}
\affiliation{Department of Physics and Astronomy, Graduate School of Science and Engineering, Kagoshima University, 1-21-35 Korimoto, Kagoshima 890-0065, Japan}

\author[0000-0003-0769-8627]{Masao Saito}
\affiliation{National Astronomical Observatory of Japan, 2-21-1 Osawa, Mitaka, Tokyo 181-8588, Japan}



\begin{abstract}

We report the Australian Telescope Compact Array and Nobeyama 45~m telescope detection of a remarkably bright ($S_\mathrm{1.1mm} = 44$~mJy) submillimeter galaxy MM~J154506.4$-$344318 in emission lines at 48.5 and 97.0~GHz, respectively. We also identify part of an emission line at $\approx 218.3$~GHz using the Atacama Large Millimeter/submillimeter Array (ALMA). Together with photometric redshift estimates and the ratio between the line and infrared luminosities, we conclude that the emission lines are most likely to be the $J = 2$--1, 4--3, and 9--8 transitions of $^{12}$CO at redshift $z = 3.753 \pm 0.001$. ALMA 1.3~mm continuum imaging reveals an arc and a spot separated by an angular distance of $1\farcs 6$, indicative of a strongly-lensed dusty star-forming galaxy with respective molecular and dust masses of $\log{M_{\rm mol}/M_\Sol} \approx 11.5$ and $\log{M_{\rm dust}/M_\Sol} \approx 9.4$ after corrected for $\approx 6.6\times$ gravitational magnification. The inferred dust-to-gas mass ratio is found to be high ($\approx 0.0083$) among coeval dusty star-forming galaxies, implying the presence of a massive, chemically-enriched reservoir of cool interstellar medium at $z \approx 4$ or 1.6~Gyr after the Big Bang.
\end{abstract}




\section{Introduction} \label{sec:intro}

%    It is becoming clear that there is an extremely-bright ($S_{\rm 1.1mm} > 40$~mJy) population of (sub)millimeter galaxies \citep[SMGs,][]{Blain02, Casey14}, via square-degree scale surveys using far-infrared (FIR) to mm single-dish telescopes (e.g., {\it Herschel}: \citealt{Eales2010}, ASTE: \citealt{Scott12}, APEX: \citealt{Weiss2009_APEX}, JCMT: \citealt{Geach2017}, SPT: \citealt{Vieira2010}). 
    Pioneering (sub)millimeter extragalactic surveys by bolometer cameras on single dish telescopes routinely discovered submillimeter galaxies \citep[SMGs,][]{Blain02, Casey14}, dusty star-forming galaxies with $L_{\rm  IR}\gtrsim10^{12} L_{\odot}$, providing a laboratory to understand extreme star formation, which cannot be seen in the local Universe. 
    Armed with next generation (sub)millimeter cameras, square-degree scale surveys using far-infrared (FIR) to mm single-dish telescopes (e.g., {\it Herschel}: \citealt{Eales2010}, ASTE: \citealt{Scott12}, APEX: \citealt{Weiss2009_APEX}, JCMT: \citealt{Geach2017,Garratt23}, SPT: \citealt{Vieira2010}) yielded a number of extremely-bright ($S_{\rm 1.1mm} \gtrsim 30$~mJy) SMGs, via early discovery and investigation of this  extremely-bright population of SMGs \citep[e.g.,][]{Swinbank10,ikarashi11,Omont11}.

 %   Thanks to their apparent high luminosity by the aid of gravitational magnification, the \emph{brightest} SMGs offer a unique opportunity to investigate their nature at the most intense peak (typical star formation rates of $\mathrm{SFR} \sim 10^3 M_\Sol$~yr$^{-1}$) of star-formation history of galaxies, through observations of cool interstellar media (ISM) traced by CO. 
    Thanks to their apparent high luminosity by the aid of gravitational magnification, the \emph{brightest} SMGs offer a unique opportunity to investigate their nature at the most intense peak (typical star formation rates of $\mathrm{SFR} \sim 10^3 M_\Sol$~yr$^{-1}$) of star formation \citep[e.g.,][]{Vieira2013,Spilker14,Spilker15,Zhang18,Litke19,Rybak20,Guruajan22}. 
    %%Their number counts (i.e., the number density of sources per unit solid angle as a function of their flux density) are expected to hold integrated information on the emergence and prevalence of massive starbursts, which is sensitive to how quickly baryonic matter collapses. Thus, the detection rate of such brightest SMGs must be the indicator to constrain galaxy formation models.  
    However, the brightest SMGs are even brighter and rarer than existing (sub)millimeter-bright sources, such as {\it Herschel}-ATLAS/SDP \citep{Negrello10}. 
    %%The number counts are therefore very uncertain because of poor statistics, and 
    It is indeed hard to spectroscopically-identify such brightest SMGs due to their optical faintness; only a couple of brightest SMGs are expected within $\sim$10~deg$^2$ \citep[e.g.,][]{Takekoshi13}.
 
    MM~J154506.4$-$344318 (hereafter, \target) was initially identified through the 4-deg$^2$ Lupus-I molecular cloud survey carried out using the AzTEC 1.1~mm camera \citep{Wilson08} attached to the ASTE 10~m telescope \citep{Ezawa04,Ezawa08}, as a $S_\mathrm{1.1mm} = 44 \pm 5$ mJy point-like object at the edge of the Lupus-I molecular cloud.  Indeed, the close proximity of this object to the molecular cloud misled us to classify it as an extremely-dense starless core shortly ($\sim 10^2$--$10^3$ yr) before protostar formation, which is only proposed theoretically \citep{Larson69, Tomida10}.
    
    Subsequent follow-up observations that we made using the Submillimeter Array (SMA), Very Large Array (VLA), ASTE, Nobeyama 45~m, Nobeyama Millimeter Array, Subaru/MOIRCS, UKIRT/WFCAM, along with archival \textit{Spitzer} (MIPS and IRAC) and \textit{Herschel} (SPIRE) data, however, suggest a gravitationally-lensed distant SMG at $z \sim 4$--5 \citep{Tamura15}. In particular, 6~cm and near-infrared (NIR) emission cannot be explained by any latest models of a starless core \citep[e.g.,][]{Tomida10, Tomida13}. 
    But a lensed SMG located far beyond the Lupus-I cloud by chance \citep{Tamura15} accounts for the multi-wavelength properties and brightness at (sub)millimeter wavelengths if a low-$z$ foreground galaxy seen in the NIR $K_S$-band (J1545B) gravitationally magnifies it.
    %%But a lensed SMG consistently accounts for the multi-wavelength properties and brightness at (sub)millimeter wavelengths if a low-$z$ foreground galaxy seen in the NIR \myrevised{$K_S$-band} (J1545B) gravitationally-magnifies a background dusty star-forming galaxy which is located far beyond the Lupus-I cloud by chance \citep{Tamura15}.
    
    If this is the case, \target{} is the brightest of $\sim 1400$ SMGs found in the AzTEC/ASTE 1.1-mm deep extragalactic survey \citep[e.g.,][]{Kohno08, Scott12} and places an interesting constraint on the physical properties of SMGs at the brightest-end of 1.1-mm extragalactic source counts. The shallow slope of the counts at $S_\mathrm{1.1mm} \sim 40$ mJy is also in favor of the presence of gravitational magnification \citep{Takekoshi13, Tamura15}.\footnote{If a certain, small fraction of galaxies at the brightest end of the number counts are gravitationally magnified, the brightest end extends toward higher flux densities, making the slope of the counts shallower at the highest flux densities.}
    
    In this paper, we report the detection of \target{} in several emission lines, which are most likely to be rotational transition lines of carbon monoxide (CO), using the Australian Telescope Compact Array (ATCA), the Nobeyama 45~m telescope, and the Atacama Large Millimeter/submillimeter Array (ALMA).  We also present {results of $0\farcs 4$ resolution} 1.3-mm continuum imaging with ALMA, suggesting the existence of gravitational magnification.  
    %%The photometric redshift obtained by our earlier work and the ratio between the line and total infrared (IR) luminosities strongly suggests that these are $^{12}$CO (2--1), (4--3), and (9--8) at redshift $z = 3.753 \pm 0.001$, which confirms the extragalactic origin of \target{}.  
    %%
    
    The rest of the paper is organized as follows.  In section 2 we describe observations carried out using the ATCA and Nobeyama 45~m telescopes.  Section 3 describes the results.  In section 4 we discuss the results and summarize the paper.
    
    Throughout this paper we use a concordance cosmology with the matter and dark energy densities $\Omega_\mathrm{m} = 0.3$, $\Omega_{\Lambda} = 0.7$, and a Hubble parameter $H_0 = 70$ km~s$^{-1}$~Mpc$^{-1}$.

\section{Observations} \label{sec:obs}

\subsection{ATCA Observations and Data Reduction}

    The ATCA observations covering 37--50 GHz were made on 2013 October 10--12 and 2014 April 4--6 (program ID: C2910).  The array configurations employed in 2013 and 2014 were H214 and H168, respectively, while we do not use the longest baselines including the CA06 antenna because of their poor atmospheric phase stability.  The resulting baseline length ranges 82--247~m and 61--192~m for the H214 and H168 configurations, respectively. The receivers were tuned to mostly cover the Q band of ATCA, which required 5 tuning setups covering 33.4--37.1, 36.9--40.6, 40.4--44.0, 43.6--47.2, and 46.8--50.4~GHz.  The lowest frequency setup, however, was flagged becase of poor weather conditions.  The Compact Array Broadband Backend (CABB) correlator was configured with the 1 MHz resolution mode for all of the receiver setups.  The instrumental delay was checked and fixed using the bright radio source 3C~279 before starting the observing tracks.  3C~279 was also used for bandpass calibration.  We observed J1451$-$375 ($S_\mathrm{7mm} \approx 1$ Jy, 10.7 deg away from the target) every 10~min for complex gain calibration.  Mars was observed once a night for absolute flux calibration.  The absolute flux accuracy is estimated to be $< 20\%$.
    
    Calibration and imaging were processed using the \textsc{Miriad} software \citep{Sault95}.  The data with poor phase stability were flagged before imaging.  The resulting integration time per tuning was $\sim 80$--90~min on average. The calibrated $uv$ data were imaged using a natural weighting, and then the dirty image was beam-deconvolved down to a $1\sigma$ level using \textsc{Miriad} tasks \texttt{invert} and \texttt{clean}, respectively.  This yields a synthesized beam size of $5 \farcs 8 \times 3\farcs 7$ (PA = $+81\arcdeg$) at 48.6~GHz. The resulting r.m.s.\ noise levels for a spectral cube with a 200 km~s$^{-1}$ resolution and a continuum image integrated over 13~GHz are 0.7--1.1 mJy~beam$^{-1}$ (depending on observed frequencies) and 0.036 mJy~beam$^{-1}$, respectively.

\subsection{Nobeyama 45~m Observations and Data Reduction}

    In order to observe an upper state transition of the emission line detected with ATCA, we carried out the 45~m observations during two periods, 2015 February 25 to March 4 (ID: CG141007) and 2016 February 26 to March 8 (program ID: CG151009).
    Because of the low declination of the source ($\delta \approx -35\arcdeg$), the elevation angle at Nobeyama ranges $\sim 15$--20 deg.  We used the two-beam dual-polarization TZ receiver \citep{Nakajima13} together with the SAM45 digital spectrometer equipped with four spectral windows, each of which has a 2048~MHz bandwidth and 4096 spectral channels \citep{Iono12}.%
    %%\footnote{SAM45 is an exact copy of ALMA's ACA correlator \citep{Kamazaki12}.} 
    The receiver and spectrometer were tuned such that the lower sideband (LSB) were centered at two slightly different frequencies{,} 96.8 and 97.0~GHz{,} in order to distinguish a broad spectral line in radio frequency (RF) from potential baseline wiggles of the intermediate frequency (IF) bandpass. The beam size (half-power beam width) is $18''$. The native velocity resolution of the spectrometer is $\approx$~1.5~km~s$^{-1}$ per channel, whereas {adjacent} 64 channels were binned to achieve $\approx$100~km~s$^{-1}$ resolution. In total, four spectral windows were configured to cover four IFs of two beams and two polarizations.
    
    During observations, we position-switched the two beams of TZ with a period of 20~s (visiting on and off positions for 5~s each, plus antenna slew time), so that one of the beams is always on source. Intensity calibration was made 2--3 times per hour by using the standard single-temperature chopper-wheel method.  Note that the atmosphere is reasonably approximated by a plane parallel slab even for such low elevation angles. The pointing was checked every 1~hr and pointing accuracy during the runs was typically $3''$ r.m.s.\ under moderate wind speeds $< 5$ m~s$^{-1}$. The focusing correction was made at the secondary mirror based on a model describing homologous deformation of the primary mirror, while no additional focusing correction was made, which potentially limited the accuracy of our aperture efficiency estimate.
    The standard calibration source M17SW was observed several times at similar elevation angles for aperture efficiency ($\eta_\mathrm{ap}$) measurements.
    By comparing the observed peak temperature of CS~(2--1) ($T_{A}^{\ast}=3.6$~K) with a past measurement of its main beam temperature ($T_\mathrm{MB}=9.0$~K), the main beam efficiency was estimated to be $\eta_\mathrm{MB} \approx 0.40$.
    If assuming a typical conversion of $\eta_\mathrm{MB} = 1.2 \eta_\mathrm{ap}$,\footnote{Assuming a Gaussian approximation for the beam solid angle and an aperture illumination with $-12$~dB edge taper \citep{Baars07}.} $\eta_\mathrm{ap} \approx 0.33$, which corresponds to a gain of $\approx 4.7$~Jy~K$^{-1}$.
    According to the 45~m status report in the 2015--2016 season,\footnote{\url{https://www.nro.nao.ac.jp/~nro45mrt/html/prop/eff/eff2015.html}} this is 80\% of the TZ aperture efficiency measured at 98~GHz at an elevation angle of $\approx 40\arcdeg$, which is consistent with but slightly lower than the past estimates, which is likely due to low elevation {angles} of our observations.
    %We note that this is consistent with the previous measurements of Saturn in 1995 showing that $\eta_\mathrm{ap}$ was about 80\% at 20 degrees relative to 40 degrees.

    Prior to the data reduction, we selected the data with good pointing accuracy under good wind speeds ($< 3$~m~s$^{-1}$) and fair system noise temperatures ($T_{\rm sys} < 200$~K), leaving the data taken on 2016 February 26--28 and March 1, 3, and 4. 
    %%Intensity calibration was made using a single-temperature chopper wheel method.
    Flagging, spectral-baselining, and coadding of the calibrated spectra were done using the ATNF Spectral Analysis Package (\textsc{asap}) shipped with \textsc{casa} \citep[version 5.0.0,][]{CASATeam22}.
    We loaded the observed data (in the format of \textsc{NewStar} \citealt{Ikeda01}) into \textsc{casa} as the \textsc{asap} Scantable format.
    We then flagged the data with the r.m.s.\ noise levels being higher than those expected from the nominal system noise temperature, where spectral baseline fluctuations likely dominate the noise levels of the instantaneous spectra.
    The resulting on-source time is 5.46~hr.
    The offset and slope of each 5~s spectrum is subtracted by fitting a linear function to the spectral baseline.
    The central $\approx 1000$~km~s$^{-1}$ where the line is expected and the $\approx 200$~km~s$^{-1}$ band edges were masked when the linear fitting is performed.
    The baseline-subtracted spectra are coadded with weighting of $T_\mathrm{sys}^{-2}$ to obtain the final spectrum.
    The resulting r.m.s.\ noise level with a 100 km~s$^{-1}$ resolution is 1.8~mJy.

%%...................................................................

\begin{table}[t]
\caption{The ATCA and Nobeyama 45m results.}
\begin{center}
    \begin{tabular}{ccc}
    \hline
    \hline
    Line                    & CO (2--1)             & CO (4--3) \\
    \hline
    $\alpha_{\rm J2000}$       & $15^{\rm h} 45^{\rm m} 6^{\rm s}.35$  & $\cdots$ \\
    $\delta_{\rm J2000}$       & $-34\arcdeg 43' 17\farcs 8$           & $\cdots$ \\
    Frequency (GHz)$^\dag$     & $48.51 \pm 0.007$        & $96.99 \pm 0.017$ \\
    $z_{\rm CO}$$^\dag$        & $3.7523 \pm 0.0006$     & $3.7537 \pm 0.0006$ \\
    FWHM (km\,s$^{-1}$)$^\dag$ & $543 \pm 106$          & $625 \pm 120$ \\
    $S\Delta V$ (Jy\,km\,s$^{-1}$)$^\ddag$
                            & $3.0 \pm 0.5$         & $6.5 \pm 0.7$\\
    $L^\prime_{\rm CO}$ ($10^{11}$ K\,km\,s$^{-1}$\,pc$^2$)$^\flat$
                            & $4.2 \pm 0.7$     & $2.3 \pm 0.2$ \\
    $M_{\rm mol}$ ($10^{12} M_\Sol$)$^{\flat\sharp}$
                            & $1.9 \cdot (\frac{0.88}{R_{21}})^{-1}$    & $1.5 \cdot (\frac{0.52}{R_{41}})^{-1}$ \\
    \hline
  \end{tabular}
\end{center}
\label{tab:result}
$^\dag$ Value and error are derived from a single-component Gaussian fit.\\
$^\ddag$ Value and error are derived from a spectral integration over a range of $\pm 3 \bar{\sigma}$ from the observed-frame frequency of CO, i.e., $\nu_{\mathrm{CO}} / (1 + \bar{z})$, where $\bar{z}$ and $\bar{\sigma}$ are obtained from the average redshift ($z=3.753$) and the average FWHM (590~km~s$^{-1}$) between the two observations.\\
$^\flat$ Values are not corrected for lensing magnification.\\ 
$^\sharp$ A CO~(1--0) to $M_{\rm mol}$ conversion factor $\alpha_{\rm CO} = 4.0~M_{\Sol}~{\rm (K~km~s^{-1}~pc^2)}^{-1}$ \citep{Dunne22} is assumed.
$R_{21}$ and $R_{41}$ are the CO(2--1)-to-CO(1--0) and CO(4--3)-to-CO(1--0) brightness temperature ratios, respectively. $R_{21} = 0.88 \pm 0.07$ and $R_{41} = 0.52 \pm 0.14$ \citep{Harrington21} are assumed throughout the paper. 
\end{table}
%%...................................................................

%%...................................................................
\begin{figure*}[!th]
    \includegraphics[width=1.0\textwidth]{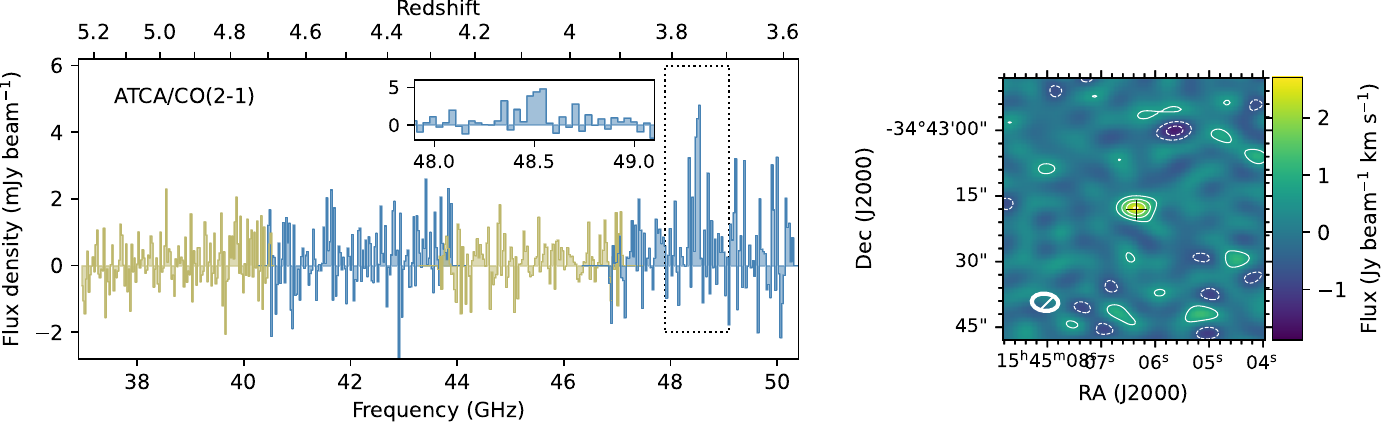}
    \caption{The ATCA spectrum and image of \target.  
    (Left) The 13~GHz wide ATCA spectrum with a 200~km~s$^{-1}$ resolution.  Four tunings are shown in different colors.  The inset highlights the line feature outlined by the dotted box.  The continuum is not subtracted.
    (Right) The CO (2--1) integrated intensity image of \target{} taken with the ATCA which shows an $8.6\sigma$ line feature at the SMA position marked with the cross \citep{Tamura15}. The contours start at $2\sigma$ with a separation of $2\sigma$. Negative values are shown in dashed contours. The synthesized beam size is indicated by an ellipse at the bottom-left corner.
    }
    \label{fig:result_atca}
\end{figure*}
%%...................................................................

\subsection{ALMA Imaging}

    We used high-angular resolution ALMA imaging data in order to search for another spectral line and to confirm the presence of {strong} lensing. As part of the Soul of Lupus with ALMA \citep[SOLA,][]{Saito15} project, the 1~mm continuum data were taken on 2016 April 1 (ID: 2015.1.00512.S; {PI.\ Itziar} de Gregorio-Monsalvo) using the Band 6 receivers in the C36--2 to --3 configurations \citep{Santamaria-Miranda21}.  The receivers were tuned at $\nu_\mathrm{LO} = 225.267$ GHz (1.3~mm). The correlators with the frequency-division mode were configured so that one of the basebands {covered} $2\times 0.469$~GHz for a local CO~(2--1) line whereas the rest {were} assigned to cover $3\times 1.875$~GHz.  The sky condition was reasonable with a precipitable water vapor (PWV) of 1.6~mm.  The complex gain was calibrated using a nearby radio-loud quasar J1534$-$3526.  The bandpass calibration was performed on J1517$-$2422, which was also used as a secondary flux calibrator ($S_\mathrm{233\,GHz} = 2.17 \pm 0.11$~Jy on 2016 March 27).  The resulting on-source time was 211~s.
    
    The $uv$ data were calibrated in a standard manner using \textsc{casa} (version 4.5.3).  The calibrated $uv$ data were then imaged using the \textsc{casa} task {\texttt{clean}} with natural weighting to make a spectral cube, resulting in a synthesized beam of $0\farcs 95 \times 0\farcs 82$ (PA = $-82\arcdeg$) {and r.m.s.\ noise level} of 0.94~mJy~beam$^{-1}$ at a spectral resolution of 15.625~MHz. 
	For continuum imaging, we employ the super-uniform weighting to achieve a high spatial resolution, which yields a synthesized beam of $0\farcs 51 \times 0\farcs 46$ (PA = $-69\arcdeg$) and the point source sensitivity of 0.14~mJy~beam$^{-1}$. The absolute flux accuracy is estimated to be 10\%.

    %continuum image and were deconvolved down to 210 $\mu$Jy~beam$^{-1}$, resulting in the spatial resolution of $0\farcs 96 \times 0\farcs 95$ (PA = $+27\arcdeg$) and the r.m.s.\ noise level of 64~$\mu$Jy~beam$^{-1}$. 

    We also retrieved public pipeline-processed band 3 and 7 images (ID: 2019.1.00245.S and 2018.1.00126.S) from the ALMA data archive for a subsequent spectral energy distribution (SED) analysis (\S~\ref{sect:intrinsic}). The beam sizes of the band 3 and 7 data are $2\farcs 7 \times 2\farcs 2$ and $0\farcs 89 \times 0\farcs 75$, respectively. The noise levels are $0.04$ and $0.12$~mJy~beam$^{-1}$, respectively. The absolute flux accuracy is estimated to be 10\%.

\section{Results}
\subsection{CO Detections and Redshift Identification}

    Across the ATCA 37--50 GHz spectrum and the integrated intensity map in Figure~\ref{fig:result_atca}, we detect a $8.6\sigma$ emission line at 48.5~GHz at a position of $(\alpha_{\rm J2000}, \delta_{\rm J2000}) = (15^{\rm h} 45^{\rm m} 6^{\rm s}.35, -34\arcdeg 43' 17\farcs 8)$, which is in agreement with the previous SMA position, ($\alpha_\mathrm{J2000}, \delta_\mathrm{J2000}) = (15^{\rm h} 45^{\rm m} 6^{\rm s}.347,\,  = -34\arcdeg 43' 18\farcs 18$) \citep{Tamura15}.  The offset between the ATCA and SMA positions is consistent with a statistical error ($0\farcs 4 \pm 0\farcs 3$), strongly suggesting that the emission feature is the counterpart to \target{}.  The velocity width and continuum-subtracted flux of \target{} are respectively $543 \pm 106$ km~s$^{-1}$ and $S\Delta V = 3.0 \pm 0.5$ Jy~km~s$^{-1}$ as listed in Table~\ref{tab:result}, which are typical among $^{12}$CO emission lines found in lensed SMGs.
    %% As reported by \citet{Tamura15}, we also detect the continuum emission at $0.210 \pm 0.035$~mJy.  The flux densities at 38.9 and 43.6~GHz are $0.12 \pm 0.05$ and $0.24 \pm 0.05$~mJy, respectively, yielding the 7~mm spectral index of $\alpha = 3.5 \pm 1.6$, where $\alpha$ is defined as $S_{\nu} \propto \nu^\alpha$.  This is consistent with dust emission with an emissivity index of $\beta_{\rm dust} = 1.4$ found with AzTEC/ASTE (1.1~mm) and the SMA \citep[1.3~mm and 890~$\micron$,][]{Tamura15}.
    As reported by \citet{Tamura15}, we also detect the continuum emission at $0.21 \pm 0.04$~mJy. The 40.4 and 46.9~GHz flux densities are $0.18 \pm 0.06$ and $0.23 \pm 0.06$~mJy, respectively, yielding the 7~mm spectral index of $\alpha \sim 1.4$, where $\alpha$ is defined as $S_{\nu} \propto \nu^\alpha$. As we will see in Section~\ref{sect:intrinsic}, we find that the obtained slope is shallower than what is expected for a single component dust emission and that the flux density exceeds the prediction from the best-fitting modified blackbody, which suggests additional contribution from a even lower $T_{\rm dust}$ component and/or free-free emission.
    %% [COMMENT FROM AKIO] Fixed the flux densities at two frequencies and the spectral index, which were obtained from newly CLEANed continuum images with CASA 6.5 by Akio Taniguchi. The previous values were something incorrect because zero-flux channels were included in the previous CLEAN images. Note that the average flux density of all SPWs (0.21 mJy) is the same as the previous value, so we do not need recalculations of other parts, i.e., SED fitting.
 
\begin{figure}[!th]
    \includegraphics[width=0.45\textwidth]{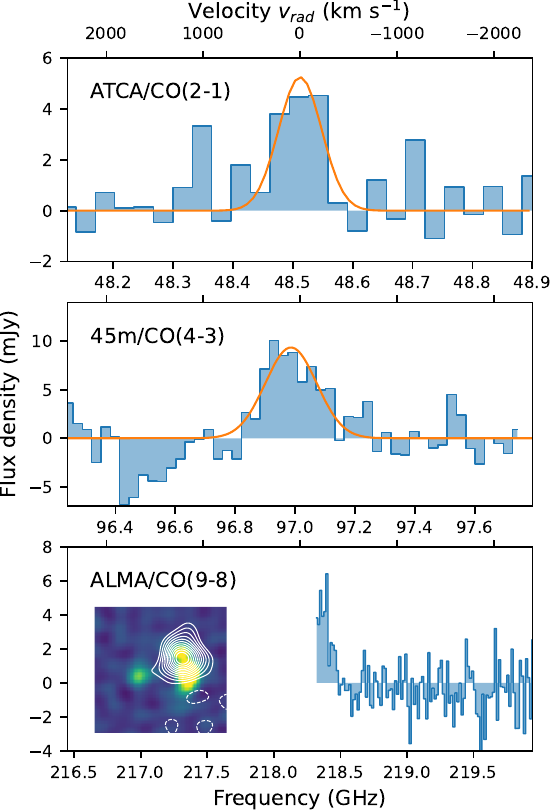}
    \caption{The CO spectrum of \target{}. From top to bottom the ATCA, Nobeyama Radio Observatory (NRO) 45~m, and ALMA band 6 spectra are shown. The orange curve shows the best-fitting Gaussian. The inset in the bottom panel shows the $4''\times 4''$ images of CO~(9--8) (contours at $-2, 2, 3, \cdots, 13\sigma$ with $\sigma=63$~mJy km s$^{-1}$) overlaid on the 1.3~mm continuum.
    }
    \label{fig:result_nro}
\end{figure}

    Figure~\ref{fig:result_nro} shows the resulting ATCA, 45~m, and ALMA spectra. The 45~m spectrum shows a broad $5.8\sigma$ emission line at 97.0~GHz and at a velocity resolution of 100~km~s$^{-1}$, with a velocity-integrated flux of $6.5 \pm 0.7$ Jy~km~s$^{-1}$. The velocity width is $625 \pm 120$ km~s$^{-1}$, which is in good agreement with that of the ATCA (Table~\ref{tab:result}). The frequency is almost exactly twice that of the ATCA emission line, which supports that they are emission lines of two rotational transitions of CO, where the upper state quantum numbers are related as $J_\mathrm{45m} = 2J_\mathrm{ATCA}$.

    Furthermore, we found a $13\sigma$ enhancement at the edge (218.3~GHz) of one of the ALMA spectral windows. The enhancement is so bright that the most likely interpretation is the blue part of another CO transition. A slight spatial-offset between the integrated intensity and 1.3~mm continuum images (see the inset of the ALMA spectrum in Figure~\ref{fig:result_nro}) is probably because the blue part of the line is associated with the northern component of the western arc, which we will see in section~\ref{sect:alma-lens}. If this is the case, the only solution that explains ATCA, 45~m, and ALMA detections is the CO (2--1), (4--3), and (9--8) transitions at the redshift of $z = 3.753 \pm 0.001$. 
 
    This is consistent with previous photometric redshift estimates derived from mid-IR-to-radio photometry and SED templates of dusty star-forming galaxies \citep[Arp~220, SMM~J2135$-$0201, and average SMG,][]{Silva98, Michalowski10, Swinbank10}; $z = 4.67^{+0.88}_{-0.74}$, $4.06^{+0.92}_{-0.11}$ and $4.20^{+0.87}_{-0.64}$, respectively \citep[all error bars represent the 68\% confidence intervals]{Tamura15}. We find only a single line over the 13~GHz bandwidth of ATCA, which also places a redshift upper limit of $z < 7$ if the line is $^{12}$CO. If the transitions are ($J = 1$--0 and 2--1) or ($J = 3$--2 and 6--5), the redshift would be $z = 1.876$ or 6.128, respectively, which are unlikely because they are well outside the photometric redshift estimates from the SED fits and cannot explain the ALMA spectrum.
 
    In order to further assess the attribution of the lines, we use the $L_\mathrm{FIR}$-to-$L'_\mathrm{CO}$ correlation \citep[e.g.,][]{Iono09} to predict a CO intensity at 48.5~GHz from the FIR luminosity of MM J1545 by following the prescription presented by \citet{Tamura14}.  If we assume the redshift of $z = 3.753$, the inferred FIR luminosity is estimated to be $\log{L_\mathrm{FIR}/L_\Sol} \sim 13.5$--14 as presented in section~\ref{sect:intrinsic}.  This yields a CO~(3--2) luminosity of $L'_\mathrm{CO} \sim 2 \times 10^{11}$ K~km~s$^{-1}$~pc$^2$, which is almost similar to the CO~(2--1) luminosity if assuming a typical luminosity ratio of $L'_\mathrm{CO(3-2)}/L'_\mathrm{CO(2-1)} \gtrsim 0.88$ found in dusty star-forming galaxies \citep[e.g.,][]{Harrington21}.  The predicted flux is then $I_\mathrm{CO(2-1)} \sim 1$--2 Jy~km~s$^{-1}$, which is consistent with that observed at 48.5~GHz with the ATCA.  This also rules out other possible lines such as $^{13}$CO, HCN, HCO$^+$ since the non-$^{12}$CO lines should be an order of magnitude fainter.  Thus, we conclude that the emission lines at 48.5, 97.0, and 218.2~GHz are attributed to three $^{12}$CO rotational transitions of $J = 2$--1, 4--3, and 9--8, respectively.

%%...................................................................
%%...................................................................

\begin{figure*}[!th]
	\begin{center}
		\includegraphics[scale=0.65,bb=0 0 491 479]{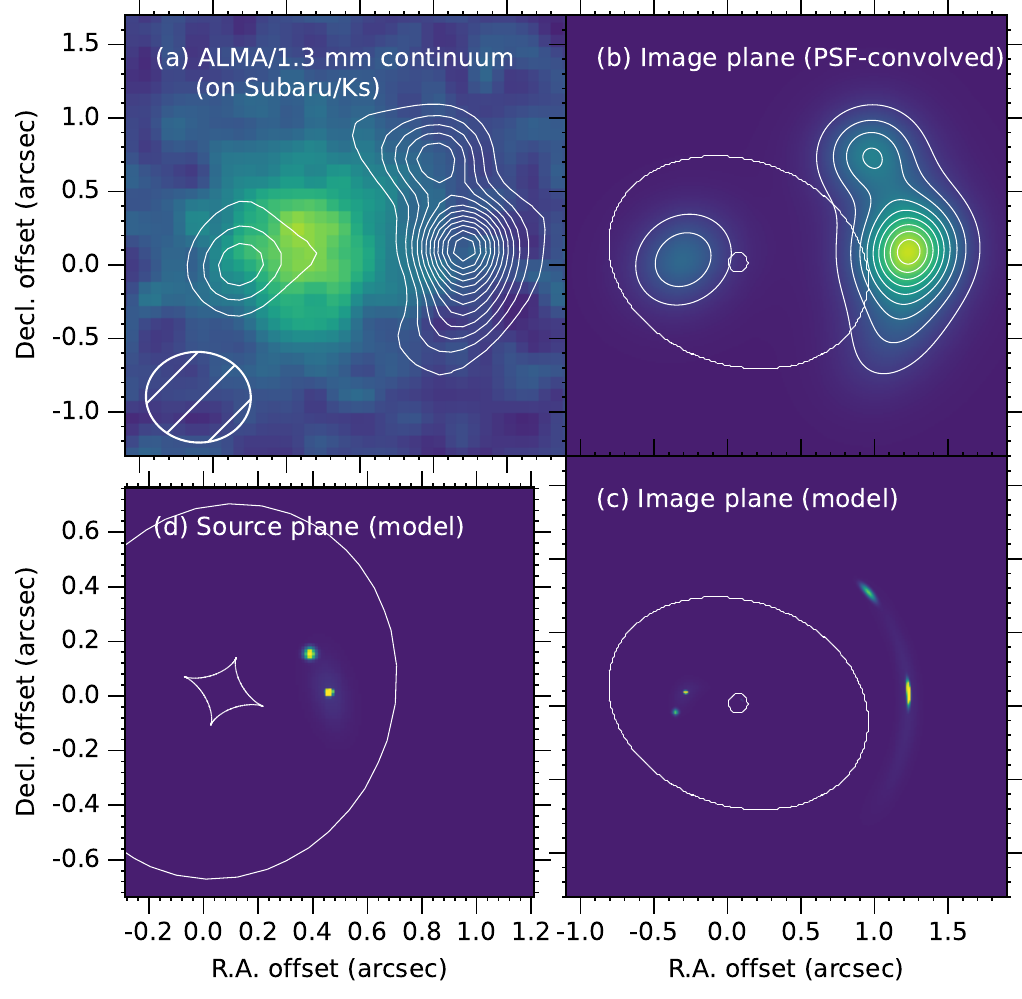}
		\caption{
			(a) The superuniform-weighted ALMA 1.3~mm continuum image {(contours) overlaid on the Subaru/MOIRCS $K_S$-band image}. The contours are drawn at $(5, 10, 15, \cdots)\times \sigma$, where $\sigma=0.14$~mJy~beam$^{-1}$ is the $1\sigma$ noise level. The beam size is indicated by the hatched ellipse at the bottom-left corner. 
            {The $K_S$-band image shows the foreground lensing galaxy J1545B \citep{Tamura15}.}
            (b) The modeled image plane brightness which is convolved by ALMA's point spread function (PSF). The contours are drawn at $10, 20, \cdots 90\%$ of the peak brightness. The large and small ellipses represent the critical curve.
            (c) The same as (b) but is deconvolved by the ALMA PSF.
            (d) The modeled source plane brightness with the caustics (solid curves).
		}\label{fig:alma}
	\end{center}
\end{figure*}

\subsection{ALMA Imaging and Gravitational Lens Model}\label{sect:alma-lens}

    As shown in Figure~\ref{fig:alma}a, the resulting ALMA continuum image reveals an arc to the west with an unresolved spot to the east, strongly suggesting the presence of strong lensing. The image shows at least two brightness peaks embedded in the western arc, suggesting two star-forming clumps embedded in this galaxy. We use the gravitational lens code, \textsc{Glafic} \citep{Oguri10}, to simply model the galaxy--galaxy lensing system.  As a mass model, we assume a cored singular-isothermal ellipsoid situated at $z = 0.5$.  The photo-$z$ of the lensing galaxy J1545B {(\citealt{Tamura15}; see the background $K_S$-band image in Figure~\ref{fig:alma}a)} is not constrained very well, while the separation of $1\farcs 6 \approx 2 \theta_\mathrm{E}$ is reproduced for $z = 0.5$ if the velocity dispersion of J1545B follows the NIR version of the Faber-Jackson relation \citep{LaBarbera10}, as discussed in \citet{Tamura15}.  We first use only the positions of the two peaks to roughly constrain the mass model. We then perform a $\chi^2$ fit to the observed brightness distribution to put more stringent constraints on the mass model and intrinsic source brightness on the source plane. Three 2-dimensional Gaussians are used for realization of the source plane brightness; two compact axisymmetric Gaussians and an extended Gaussian to represent two brightness peaks and an extended diffuse component, respectively.  Before applying the $\chi^2$ fits we smooth the modeled image plane with the clean beam of $0\farcs 51 \times 0\farcs 45$ (PA = $-69\arcdeg$). We do not consider a model fit in the visibility domain for simplicity. We do not take an external shear into account because the degree of freedom is small. A noise level of $\sigma = 0.15$ mJy~beam$^{-1}$ is used for $\chi^2$ calculation.
 
    The results are shown in Figure~\ref{fig:alma}b, \ref{fig:alma}c, and \ref{fig:alma}d for the image and source planes. {The centroid of J1545B on the $K_S$-band image (Figure~\ref{fig:alma}a) is in good agreement with the predicted position of the mass model.}  The velocity dispersion of the lens (200~km~s$^{-1}$) is consistent with that estimated from the $K_S$-band version of the Faber--Jackson relation \citep{LaBarbera10}. The image plane reproduces the western arc with two clumps and the eastern spot although the extended emission at low brightness levels is not reproduced very well. On the source plane we find an interesting spatial structure that shows two bright spots embedded in an extended disk while high-resolution imaging is necessary to confirm the detailed internal structure of \target{}. The inferred magnification factor is $\mu_{\rm g} = 6.6$ although this may have a large uncertainty depending on the extent of source plane brightness.

    \begin{figure}[!th]
        \includegraphics[width=0.40\textwidth]{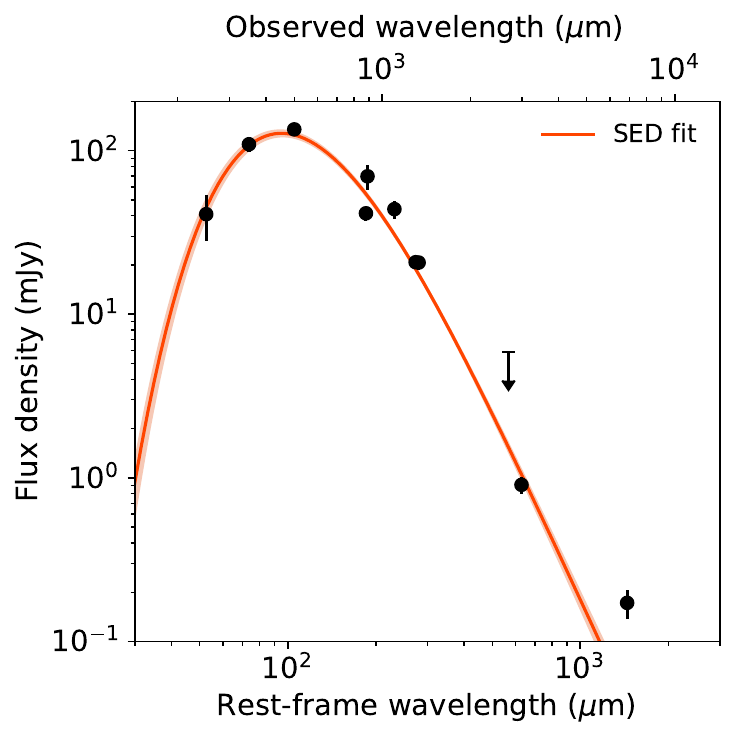}
        \includegraphics[width=0.45\textwidth]{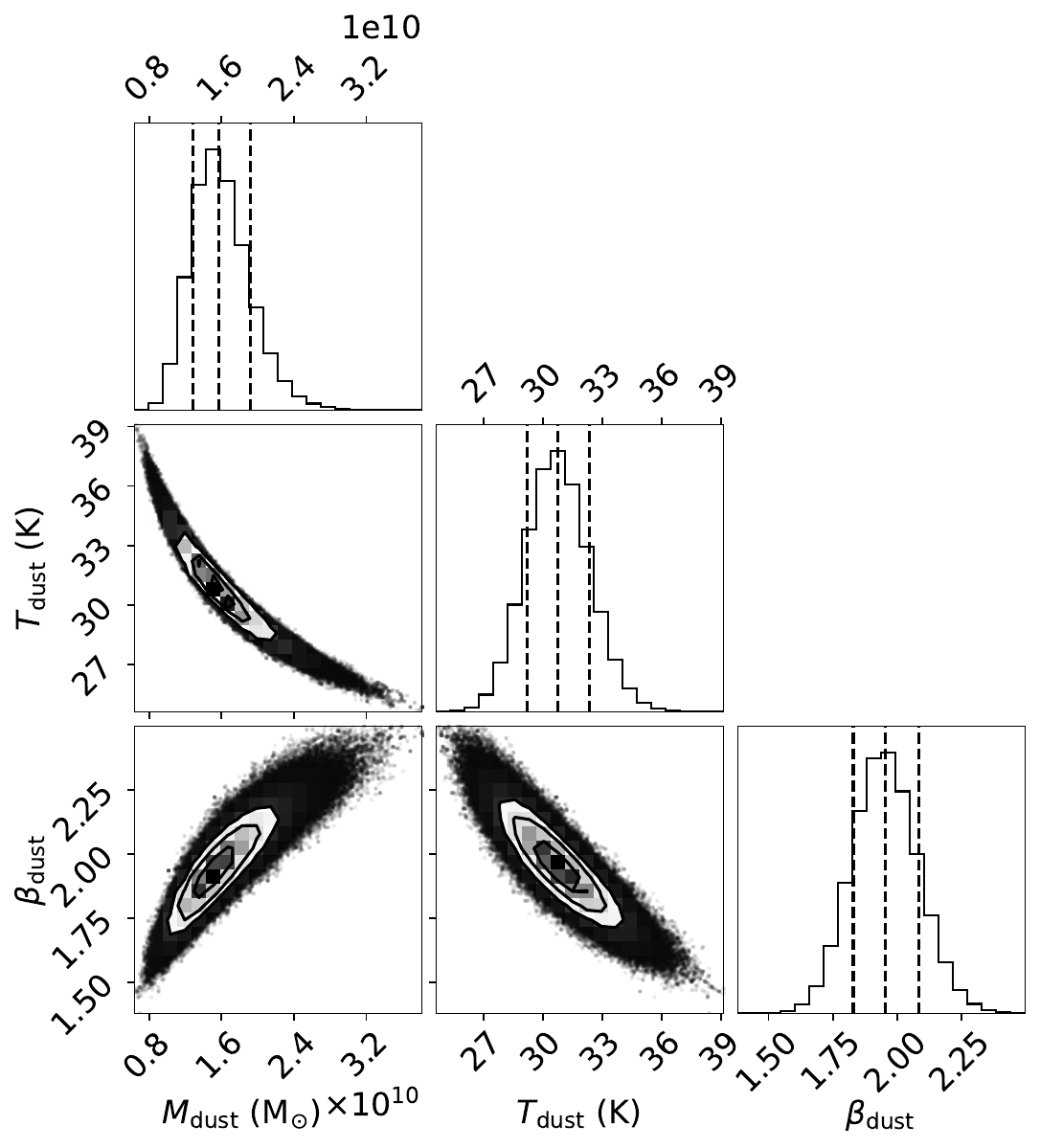}
        \caption{Top: The FIR-to-mm spectral energy distribution (SED) of \target{} (the red curve). The light-red band associated with the best-fitting SED represents the 68\% confidence interval.
        Bottom: The posterior distribution function (PDF) of the parameters in the SED fits. The contours in the 2-dimensional PDF are drawn at 1, 1.5, and $2\sigma$. The vertical dashed lines in the marginalized 1-dimensional PDF are shown at 16, 50, and 84 percentile. {The flux densities and the dust mass are not corrected for lensing magnification.} See \S~\ref{sect:intrinsic} for details.
        }
        \label{fig:sed}
    \end{figure}

\section{{Discussions}}

\subsection{{Validity of the Magnification Factor}}
    %\textbf{[To-do] What should we discuss about? (1) Revisiting gravitational lensing using the SED templates, L'co-dV relation and intrinsic properties, (2) excitation state of molecular gas}
    
    The ALMA image clearly reveals multiple images split by the strong gravitational lensing effect while the angular resolution is still insufficient for full characterization of the lensing system, which limits the accuracy of the magnification factor.  Instead, an excess of an observed CO luminosity from the correlation between (unlensed) CO~(1--0) luminosity and velocity dispersion places an independent constraint on a magnification factor \citep{Harris12}.  This empirical luminosity--line width relation is similar to the Tully--Fisher relation \citep{Tully77} relating the luminosity and rotation velocity of spiral galaxies and has been characterized for CO emission of the SMG population \citep{Bothwell13, Harris12} as
    \begin{equation} \label{eq:lv}
        L_\mathrm{CO}^{\prime\ 11} = (\Delta V_\mathrm{400})^{1.7}/3.5,
    \end{equation}
    where $L_\mathrm{CO}^{\prime\ 11}$ is (unlensed) luminosities of CO~(1--0) in units of $10^{11}~\mathrm{K~km~s^{-1}~pc^2}$ and $\Delta V_\mathrm{400}$ is the FWHM line width in units of $400~\mathrm{km~s^{-1}}$.  The $L'_\mathrm{CO}$--$\Delta V$ correlation of unlensed SMGs is known to be scattered because of unknown inclination angles of the SMGs, but this is still useful to roughly constrain the magnification factors of lensed SMGs.
 
    As we {will see in the next section}, the molecular masses derived from CO~(2--1) and (4--3) with their conversion factors, $R_{21}$ and $R_{41}$, typically found among SMGs \citep[e.g.,][]{Harrington21, Hagimoto23} are almost the same, suggesting that the inferred CO spectral line energy distribution of \target{} is similar to those found in SMGs. Thus, it is reasonable to assume $L_\mathrm{CO}^{\prime\ 11} = 3.7 R_{21}^{-1} \approx 4.2$ derived from CO (2--1) for the CO (1--0) luminosity. Eq.~\ref{eq:lv} predicts the CO (1--0) luminosity of $L_\mathrm{CO}^{\prime\ 11} = ({543}/400)^{1.7}/3.5 \approx {0.48}$, which requires $\approx 9 \times$ magnification to reach the apparent CO~(1--0) luminosity. Despite the $\sim 0.5$ dex scatter in the $L'_\mathrm{CO}$--$\Delta V$ correlation, this is in reasonable agreement with the magnification factor we derived in section~\ref{sect:alma-lens}. 
    {Hereafter we assume the magnification factor to be $\mu_{\rm g} = 6.6$.}

\subsection{Intrinsic Properties}\label{sect:intrinsic}

    \begin{figure}[!th]
        \includegraphics[scale=0.90,bb=0 0 271 367]{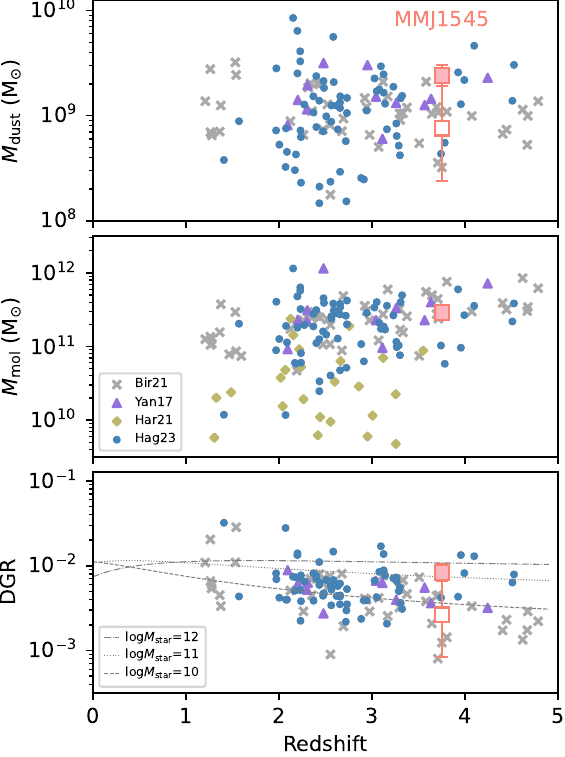}
		\caption{
			{The dust mass, molecular gas mass, and the dust-to-gas mass ratio (DGR) of \target{} (the red filled square) in comparison with lensed \citep[denoted as Bir21]{Birkin21} and unlensed \citep[denoted as Yan17, Har21, Hag23, respectively]{Yang17, Harrington21, Hagimoto23} submillimeter-bright galaxies as a function of redshift. The masses are corrected for lensing. We assume $\alpha_{\rm CO} = 4.0~M_{\Sol}~{\rm (K~km~s^{-1}~pc^2)}^{-1}$ and the same CO excitation ladder as \citet{Harrington21} for a fair comparison.
            The red open square indicates the dust mass and DGR of \target{} with an assumption of a possible lower dust emissivity ($\kappa_{\rm \star}$ = 0.11\,cm$^2$~g$^{-1}$, see \S~\ref{sect:intrinsic}). The dashed, dotted, and dash-dotted curves represent predicted redshift evolution of DGRs with stellar masses of $\log{M_\mathrm{star}/M_\Sol} = 10$, 11, 12, respectively \citep{Genzel15, Tacconi18}.}
		}\label{fig:dgr}
    \end{figure}

    %% - dust SED
    Figure~\ref{fig:sed} shows the rest-frame FIR-to-mm SED of \target{}. In addition to photometry from the literature \citep{Tamura15}, archival ALMA band 3 and 7 data are used. The aperture sizes we used for ALMA photometry are $3\farcs 5$ (band 3) and $2\farcs 0$ (band 6 and 7). The modified blackbody fits yield the dust temperature $T_\mathrm{dust} = 30.7_{-1.5}^{+1.7}$~K, the emissivity index $\beta_{\rm dust} = 1.95_{-0.12}^{+0.13}$, and the apparent IR luminosity of $\log L_\mathrm{IR}/L_\Sol = 13.55\pm 0.32$ if we take into account the heating from the cosmic microwave background ($T_\mathrm{CMB} = 12.9$~K at $z=3.753$) following the expression by \citet{daCunha13}.
    The emissivity index is higher than that previously obtained \citep[$\beta_{\rm dust} = 1.4$,][]{Tamura15} because the new ALMA 3~mm photometry makes the Rayleigh--Jeans slope steeper. Although the origin of the flux excess at 7~mm is unknown, there could be an additional component such as a lower $T_\mathrm{dust}$ component, which can be confirmed by ALMA band 1 observations.

    %% - delensed dust mass
    If we take the magnification factor of $\mu_{\rm g} = 6.6$, the delensed IR luminosity ($\log L_\mathrm{IR}/L_\Sol = 12.73\pm 0.32$) and corresponding star-formation rate {(}${\rm SFR} = 804^{+876}_{-419}~M_\sun$~yr$^{-1}$ using the \citet{Kennicutt12} conversion with the \citet{Kroupa93} IMF{)} are similar to those of unlensed SMGs which are in the bright-end ($S_{\rm 850\mu m} \approx 10$~mJy) of their number counts \citep[e.g.,][]{Geach2017, Hatsukade18, Simpson20, Fujimoto23}. 
    %%
    % log10(M_\Sol/(erg/s))=33.585009 -> 33.585 + 13.55 = 47.135
    % log10(mu_g=6.6) = 0.819544
    % -> 47.135-0.819544-43.41=2.9054 (+/-0.32) -> 10^2.9054 = 804 (+876/-419) M_\Sol/yr

    Remarkably, the dust temperature found in \target{} is relatively low among the SMGs with a similar unlensed $L_{\rm IR}$ \citep[e.g.,][]{Reuter2020}, indicating a massive dust mass. If we approximate the dust mass absorption coefficient ($\kappa_{\rm \nu}$) as $\kappa_{\rm \star} \left( \nu / \nu_{\star}\right)^{\beta_{\rm dust}}$, with ($\kappa_{\rm \star}, \nu_{\star})$ as (10.41\,cm$^2$~g$^{-1}$, 1900\,GHz) from \cite{Draine03}, 
    the {lensed and delensed dust masses are} inferred to be {$\log M_{\rm dust}/M_{\odot} = 10.2 \pm 0.1$ and $9.4 \pm 0.1$, respectively. As shown in the top panel of Figure~\ref{fig:dgr}, the intrinsic dust mass of \target{} is large even compared with those of the largest dust reservoirs at $z \sim 4$, such as lensed SMGs from the SPT \citep{Yang17} and \textit{Herschel} surveys \citep{Hagimoto23} as well as unlensed SMGs \citep{Birkin21} from AS2COSMOS \citep{Simpson20}, AS2UDS \citep{Stach19}, and ALESS \citep{Hodge13}.}
    This could, however, have a systematic uncertainty depending on choice of the dust mass absorption coefficient. The $\kappa_{\rm \star}$ value often employed at 850~$\mu$m (i.e., $\nu_{\star} = 353$~GHz) ranges from 0.04~cm$^2$~g$^{-1}$ \citep{Draine84} to 0.3~cm$^2$~g$^{-1}$ \citep{Mathis89} with an intermediate value 0.11\,cm$^2$~g$^{-1}$ \citep{Hildebrand83}, which gives $\kappa_{\rm \star} \approx 1$--8~cm$^2$~g$^{-1}$ with an intermediate value $\approx 3$~cm$^2$~g$^{-1}$ at $\nu_{\star} = 1900$~GHz if assuming $\beta_{\rm dust} = 1.95$. In this case the dust mass may be reduced by $\sim 0.5$~dex, yielding {an intrinsic dust mass of $\log M_{\rm dust}/M_{\odot} = 8.9 \pm 0.5$}, where the error includes the systematic uncertainty arising from the choice of $\kappa_{\rm \star}$.
    %% ... dust mass estimates with the older k_d values -> can be used in DGR discussions below.
    %% (353/1900)^1.95=0.03754837561 ~ 0.03755
    
    %% - delensed molecular mass
    Also, the intrinsic molecular mass is estimated to be large.  The apparent molecular masses estimated from CO (2--1) and (4--3) are $M_{\rm mol} \approx 1.9 \times 10^{12} M_\Sol$ and $1.5 \times 10^{12} M_\Sol$ if we assume CO(2--1)-to-CO(1--0) and CO(4--3)-to-CO(1--0) brightness temperature ratios found in \textit{Planck}-selected SMGs ($R_{21} = 0.88 \pm 0.07$ and $R_{41} = 0.52 \pm 0.14$, respectively; \citealt{Harrington21}) and a CO~(1--0) to $M_{\rm mol}$ conversion factor $\alpha_{\rm CO} = 4.0~M_{\Sol}~{\rm (K~km~s^{-1}~pc^2)}^{-1}$ \citep{Dunne22}, respectively (Table~\ref{tab:result}). This indicates an intrinsic molecular mass of $\log{M_{\rm mol}/M_\Sol} \approx 11.5$ although it could decrease by $\sim 0.5$~dex depending on the choice of $\alpha_{\rm CO}$.
    {This is comparable to the molecular masses found in the SMG population at $z \sim 4$, as shown in the middle panel of Figure~\ref{fig:dgr}.}

    Along with the intrinsic SFR, the molecular gas depletion time-scale is estimated to be $\sim 0.4$~Gyr. This appears to be greater than those found in the SPT sources at $z \sim 4$ \citep{Reuter2020} whereas more similar to coeval ($z \sim 4$) main-sequence galaxies \citep{Saintonge13}. This trend is consistent with that found in the Schmidt--Kennicutt relation of dusty star-forming galaxies \citep{Hagimoto23}, where 71 \textit{Herschel}-selected bright SMGs were compared with the SPT and main-sequence galaxies. This implies that star formation in \target{} is not very bursty, which is consistent with the fact that the dust temperature is relatively low.

    The {intrinsic} dust and molecular masses ({$\log M_{\rm dust}/M_\Sol \approx 9.4$} and $\log{M_{\rm mol}/M_\Sol} \approx 11.5$) give the inferred dust-to-gas mass ratio (DGR) of {${\rm DGR} \approx 0.0083$}. 
    %% 以下、全面的に改訂。
    This value is 
    {even higher than that of the Milky Way and is among the highest end of the DGR distribution found in} bright SPT and \textit{Herschel} sources (\citealt{Yang17, Hagimoto23}, see also \citealt{Peroux20})%
    {, as shown in the bottom panel of Figure~\ref{fig:dgr}}. 
    The inferred sum of gas and solid-phase metallicity would be 
    {at least $12+\log{\rm (O/H)} \gtrsim 8.5$ or $Z \gtrsim 0.6 Z_\Sol$ following the relation of \citet{Peroux20} and is likely to be $12+\log{\rm (O/H)} \gtrsim 8.9$ or $Z \gtrsim 1 Z_\Sol$ if the dust-to-metal mass ratio is as high as $\lesssim 0.4$ \citep{Peroux20}. The high DGR or metallicity at $z \sim 4$ is indicative of the presence of an underlying massive stellar component of \target{}, which is not seen in the current optical and NIR images. The curves in the bottom panel of Figure~\ref{fig:dgr} show the DGRs expected for the stellar components with the stellar mass of $\log{M_{\rm star}/M_\Sol} = 10$, 11, and 12. This suggests that \target{} should have a stellar component with at least $\gtrsim 10^{10} M_\Sol$ and perhaps $\sim 10^{11} M_\Sol$.}
    {We note that} the DGR may be smaller by $\sim 0.5$~dex {if} the smaller value is allowed for dust mass as we saw {before}. If we take $\log M_{\rm dust}/M_{\odot} \approx {8.9}$, then the DGR is estimated to be ${\rm DGR} \approx {0.0027}$, which {is comparable to the typical values found in lensed and unlensed SMGs or DSFGs at $z \sim 4$} but still indicates {the highly-enriched ISM compared to coeval galaxies in general}.
    
    %%In summary, it is likely that \target{} has a massive, chemically-enriched reservoir of cool ISM at $z = 3.75$ or 1.6~Gyr after the Big Bang%
    %%{, though this work only adds one additional source to existing samples and thus does not provide stringent constraints on the general properties of the extremely-bright population of dusty star-forming galaxies due to a small sample size. Expanding the extremely-bright samples at $z \gtrsim 4$ will address the earlier chemical enrichment more thoroughly. Future (sub-)mm facilities, such as TolTEC on the Large Millimeter Telescope \citep{Wilson20}, will allow us to investigate this further.}

\section{{Conclusions}}

    We report detections of two emission lines at 48.5 and 97.0 GHz in a $S_\mathrm{1.1mm} = 44$~mJy SMG, MM~J154506.4$-$344318 (\target), using the ATCA and the Nobeyama 45~m telescope. We also find the blueshifted part of an emission line at $\approx 218.3$~GHz using archival ALMA data. Together with the photometric redshift estimates and the ratio between the line and IR luminosities, we conclude that they are most likely to be the (2--1), (4--3), and (9--8) rotational transitions of $^{12}$CO at redshift $z = 3.753 \pm 0.001$. The ALMA continuum imaging confirms \target{} to be a gravitationally-lensed SMG with a magnification factor of $\mu_\mathrm{g} \approx 7$, suggesting the presence of an intrinsically massive, chemically-enriched reservoir of cool ISM
    {at $z = 3.75$ or 1.6~Gyr after the Big Bang}.

    If assuming a nominal magnification of $\mu_\mathrm{g} = {6.6}$ for \target{}, the intrinsic {dust} and molecular mass{es} are still high even after correcting for magnification ({$\log{M_\mathrm{dust}/M_\Sol} \approx 9.4$ and $\log{M_\mathrm{mol}/M_\Sol} \approx 11.5$}), 
    {which is indicative of a high DGR ($\approx 0.0083$)}. 
    Such a starburst galaxy {with a chemically-enriched reservoir of cool ISM} at $z \sim 4$ is rare even compared with existing samples and will provide a unique opportunity to investigate spatially-resolved properties of ISM and star-formation in the early universe. 
    {This work, however, only adds one additional source to existing samples and thus does not provide stringent constraints on the general properties of the extremely-bright population of dusty star-forming galaxies due to a small sample size. Expanding the extremely-bright samples at $z \gtrsim 4$ will address the earlier chemical enrichment more thoroughly. Future (sub-)mm facilities for wide-field imaging \citep[e.g., TolTEC,][]{Wilson20} and wideband spectroscopy \citep[e.g., DESHIMA 2.0, FINER,][]{Taniguchi22, Rybak22, Tamura24} will allow us to investigate this further. Also,}
    stellar and nebular properties in the rest-frame optical are also necessary for comprehensive understanding. Higher-resolution ALMA and \textit{James Webb Space Telescope} imaging will play an important role in fully characterizing the physical properties of \target{}.

    %%

%% \section{Appendix: Custom Pipeline for Nobeyama 45 m Reduction}
%%

\section{Acknowledgement}
{We acknowledge the referee for fruitful comments.}
We thank Takuma Izumi for fruitful suggestions on the ATCA observations and Yuki Yamaguchi and Ryo Ando for their support on the Nobeyama 45~m observations.
This work is supported by KAKENHI (No.\ 15H02073, 19K03937, 20001003, 22H04939, 22KJ1598, 23K20035) and NAOJ ALMA Scientific Research grant No.\ 2018-09B and {NAOJ-ALMA-321}. IdG acknowledges support from grant PID2020-114461GB-I00, funded by MCIN/AEI/10.13039/501100011033. The Australia Telescope Compact Array is part of the Australia Telescope National Facility, which is funded by the Commonwealth of Australia for operation as a National Facility managed by CSIRO.  The 45 m radio telescope is operated by Nobeyama Radio Observatory, a branch of National Astronomical Observatory of Japan.  This paper makes use of the following ALMA data: ADS/JAO.ALMA\# 2015.1.00512.S, 2018.1.00126.S, and 2019.1.00245.S. ALMA is a partnership of ESO (representing its member states), NSF (USA) and NINS (Japan), together with NRC (Canada), NSC and ASIAA (Taiwan), and KASI (Republic of Korea), in cooperation with the Republic of Chile. The Joint ALMA Observatory is operated by ESO, AUI/NRAO and NAOJ. Data analysis was in part carried out on the Multi-wavelength Data Analysis System operated by the Astronomy Data Center (ADC), National Astronomical Observatory of Japan. 

\software{
    Astropy \citep{AstropyColab13,AstropyColab18,AstropyColab22},
    CASA \citep{CASATeam22},
    glafic \citep{Oguri10},
    lmfit \citep{Newville23},
    matplotlib \citep{Hunter07},
    NumPy \citep{Harris20},
    pandas \citep{McKinney10,PandasTeam20},
    xarray \citep{Hoyer17},
}




\bibliography{sample631}{}
\bibliographystyle{aasjournal}



\end{document}